\documentclass[preprint,prd,aps,nofootinbib]{revtex4}

\usepackage{graphicx}

\begin{document}

\title {\large Double sneutrino inflation and its phenomenologies  }

\author{ Xiao-Jun Bi }
\email[Email: ]{bixj@mail.ihep.ac.cn}
\affiliation{ Institute of High Energy Physics, Chinese Academy of
Sciences, P.O. Box 918-4, Beijing 100039, People's Republic of China}

\author{ Bo Feng }
\email[Email: ]{fengbo@mail.ihep.ac.cn}
\affiliation{ Institute of High Energy Physics, Chinese Academy of
Sciences, P.O. Box 918-4, Beijing 100039, People's Republic of China}

\author{ Xinmin Zhang }
\email[Email: ]{xmzhang@mail.ihep.ac.cn}
\affiliation{ Institute of High Energy Physics, Chinese Academy of
Sciences, P.O. Box 918-4, Beijing 100039, People's Republic of China}

\date{\today}

\begin{abstract}

In this paper we study double scalar neutrino inflation in the minimal
supersymmetric seesaw model in light of WMAP.
Inflation in this model is firstly driven by
the heavier sneutrino field $\tilde{N}_2$ and then the lighter
field $\tilde{N}_1$. we will show that
with the mass ratio $6\lesssim M_2/M_1 \lesssim 10$ 
the model predicts a suppressed primordial scalar
spectrum around the largest scales and the predicted CMB TT quadrupole is
much better suppressed than the single sneutrino model.
So this model is more favored than the
single sneutrino inflation model. We then
consider the implications of the model on the reheating
temperature, leptogenesis and lepton flavor violation.
Our results show that the  seesaw parameters
are constrained strongly by
the reheating temperature, together with the requirement by a
successful inflation. The mixing between the first
generation and the other two generations in the right-handed
neutrino sector is tiny.
The rates of lepton flavor violating processes in our scenario
depend on only 4 unknown
seesaw parameters through a 'reduced' seesaw formula,
besides $U_{e3}$ and the supersymmetric parameters. We find that the
branching ratio of $\mu \rightarrow e \gamma$ is generally near the
present experimental limit, while ${\rm Br}( \tau \rightarrow \mu \gamma)$ 
is around $\mathcal{O}( 10^{-10} - 10^{-9} )$.

\end{abstract}

\maketitle

\section {introduction}

It is widely accepted today that the early universe has
experienced an era of accelerated expansion known as inflation
\cite{guth}. Inflationary universe has solved many problems of the
standard hot big-bang cosmology, such as the flatness and horizon
problems. In addition, it provides a causal interpretation for
the origin of the density fluctuations in the Cosmic Microwave
Background (CMB) and large scale structure (LSS).

Among current inflation models, sneutrino chaotic
inflation\cite{Yanagida0,Yanagida} is one of the promising physical
candidates where inflation is driven by the superpartner of the
right-handed (RH) neutrino. In this scenario, no extra inflaton
scalar field is needed, besides the RH sneutrinos, which are necessary
to explain the tiny neutrino mass\cite{neu} in the minimal supersymmetric
seesaw mechanism\cite{seesaw}.
Baryon number asymmetry via leptogenesis\cite{lepto}
can also be easily realized in this framework.

The single sneutrino inflation model predicts a near scale invariant
primordial power spectrum. Despite the fact that the scale invariant
primordial spectrum is consistent with current Wilkinson Microwave
Anisotropy Probe (WMAP) observations \cite{Bennett}, it is noted
that there might be possible discrepancies between predictions and
observations on the largest and smallest scales. WMAP data show a
low TT quadrupole \cite{Hinshaw} as previously detected by
COBE\cite{cobe}. In Ref.\cite{Peiris} Peiris et al. find that WMAP
data alone favor a large running of the spectral index from  blue
to red at $\gtrsim 1.5\sigma$ with $dn_S/d \ln k = -
0.077^{+0.050}_{-0.052}$. When adding LSS data of 2DFGRS\cite{2df}
the running is more favored with $dn_S/d \ln k = -
0.075^{+0.044}_{-0.045}$.

The most proper way to get the shape of the
spectrum from observations should be the primordial spectrum
reconstruction\cite{Wangyun,Wang,Lewis}.
A detailed reconstruction of the power
spectrum by Mukherjee and Wang\cite{Wang} shows that a running of the
index is favored. Ref.\cite{Lewis}
reconstructs the primordial spectrum with WMAP data and the shape
of the matter power spectrum from 2DFGRS\cite{2df}. The authors
attribute the need for the running to the first three CMB multipoles
$l=2,3,4$. They introduce power-law spectrum with a cut at large
scales and find a non-vanishing cutoff is favored at $\gtrsim
1.5\sigma$.

The statistical level of the low CMB multipoles has
been discussed widely\cite{Spergel,small00} and many models have
been built to achieve the suppressed CMB
multipoles\cite{small01,fengbo1,fengbo2}.
Although the confidence level of spectral index running is
not very high, if stands, it would severely constrain inflation model
buildings\cite{wangxl,fengbo1,running} and the single field sneutrino
chaotic inflation model would be in great challenge\footnote{ Several
authors in the literature have fitted WMAP using different codes or
adding various CMB and LSS data, they give consistent
results\cite{Lewis,fitWMAP} but with less hints for running of the
spectral index.}.

Recently we have considered a double inflation
model\cite{fengbo2,double}:
\begin{equation}\label{potential} V(\phi_1,\phi_2)={1\over
2}m_1^2\phi_1^2 + {1 \over 2} m_2^2 \phi_2^2~,
\end{equation}
where inflation is driven firstly by the heavier inflaton
$\phi_2$, then the lighter field $\phi_1$. But there is no
interruption in between. This model solves the problems of
flatness \textit {etc.} and generates a primordial spectrum
suppressed at certain small $k$ values. The CMB quadrupole
predicted can be much lower than the standard power-law
$\Lambda$CDM model. Recently, it is shown by
Kamionkowski et al.\cite{Kamionkowski} that the cross-correlation between the
CMB and an all-sky cosmic-shear map will be enhanced by such a
primordial spectrum, and this may be observable at
$2-3\sigma$\cite{Kamionkowski2}. The suppressed CMB multipoles can
also lead to many other observable consequences\cite{small02}.

In the present work, we consider the case that the two inflaton
fields consist of the two lighter sneutrinos, $\tilde{N}_1$ and
$\tilde{N}_2$ in the minimal supersymmetric seesaw model, 
while the heaviest one, $\tilde{N}_3$, does not
contribute to inflation. By fitting the resulted primordial
spectrum to the WMAP data in the next section, we get the preferred two
sneutrino masses, $M_1$ and $M_2$. We find that the double
sneutrino model is more favored than the single sneutrino model at
about 1.5$\sigma$ level. In section III, we first present our
parameterization of the seesaw model and then analyze the
implications of this model on the reheating temperature,
leptogenesis and lepton flavor violation, \textit{etc}. We find
the reheating temperature, constrained by the gravitino
problem\cite{gravitino} to be below $\mathcal{O}(10^{10} GeV)$,
gives very strong constraint on the seesaw parameter space and
our analysis is greatly simplified then. Different from a random sampling
on the 9-dimensional unknown seesaw parameter space in Ref.
\cite{Yanagida}, we can show the seesaw parameter dependence of
the predicted lepton flavor violating rate explicitly. Our analysis
shows that there is no direct connection between leptogenesis and
LFV in this model. Non-thermal leptogenesis is easily to be
achieved via the sneutrino inflaton decay. Only hierarchical
neutrino mass spectrum at low energy can be produced and the
neutrinoless double beta decay\cite{0nubeta} can not be explained
by the effective Majorana neutrino mass in the model.

%In previous sneutrino inflation papers only the lightest heavy
%singlet sneutrino is assumed to drive inflation, the resulting
%spectrum is the same as in the standard chaotic inflation. In this
%paper we generalize the model into double inflation case. There
%are three sneutrinos, we assume two of them to be degenerated, or
%the mass of the heaviest sneutrino is much larger than the other
%two. We study in this paper its consequences on CMB, reheating and
%leptogenesis.

\section{Double chaotic sneutrino inflation }

The evolution of the background fields for double sneutrino
inflation is described by the Klein-Gordon equation\footnote{To be
consistent with the usual convention, in this section,
we use $\phi_I$ to represent
the inflatons, the sneutrinos here, instead of the symbol
$\tilde{N}_I$.}:
\begin{equation}
\label{eq:KG} \ddot{\phi}_I + 3H\dot{\phi}_I + V_{\phi_I}= 0~,
\end{equation} and the Friedmann equation:
\begin{equation}
H^2=(\frac{\dot a}{a})^2 =\frac{8\pi G}{3} \left[
\frac{1}{2}\dot\phi_1^{~2}+ \frac{1}{2}\dot\phi_2^{~2} +V
\right]\,, \label{eq:hubble}
\end{equation}
where $I=1,2$, $a$ is the scale factor, the dot stands for time
derivative and $V_{x} = {\partial V}/{\partial x}$. Defining the
adiabatic field $\sigma$ and its perturbation as \cite{Gordon}:
\begin{eqnarray} \dot\sigma &=&(\cos\theta) \dot\phi_1 + (\sin\theta)
\dot\phi_2
\,,\nonumber\\
\delta\sigma &=& (\cos\theta) \delta\phi_1 + (\sin\theta)
\delta\phi_2\,, \end{eqnarray} with
\begin{equation}
  \label{eq:cos sin}
\cos\theta = -\frac{\dot{\phi_2}}{\sqrt{\dot{\phi_1}^2 +
\dot{\phi_2}^2}}\, \ , \quad \sin\theta =
-\frac{\dot{\phi_1}}{\sqrt{\dot{\phi_1}^2 + \dot{\phi_2}^2}}\, \ .
\end{equation}
The background equations (\ref{eq:KG}) and (\ref{eq:hubble})
become \begin{eqnarray} H^2 &=&\frac{8\pi G}{3} (\frac{1}{2}\dot\sigma^{~2}+V
)\,,\nonumber\\
\ddot{\sigma} &+& 3H\dot{\sigma} + V_\sigma = 0\,, \end{eqnarray} where
$V_\sigma=(\cos \theta) V_{\phi_1} + (\sin\theta) V_{\phi_2}$. We
assume an adiabatic initial condition between the perturbations $\delta
\phi_1$ and $\delta \phi_2$:
\begin{equation}
\label{eq:adiabatic}
 {\delta\phi_1\over\dot{\phi_1}}={\delta
\phi_2\over\dot{\phi_2}}~.
\end{equation}
As shown in Ref.\cite{Gordon}, if the initial perturbation is
adiabatic, it will remain adiabatic on large scales during
inflation. In this sense, inflation is equivalently driven by a
single inflaton $\sigma$ with the effective potential
$V(\sigma)= V(\phi_1)+ V(\phi_2)$.
The basic picture of inflation and perturbation in our model
is: the heavy inflaton $\phi_2$ rolls
slowly down its potential and starts to oscillate when the Hubble
expansion rate is around its mass $H\sim M_2$, while $\phi_1$ remains
slow rolling and $V(\phi_1)$ comes to dominate the inflaton energy
density. Hence, inflation is not suspended during the transition.

The effective potential $V(\sigma)$, as
well as the background evolution, is determined by the initial
values of $\phi_1$, $\phi_2$ (i.e. $\phi_{1i}$ and $\phi_{2i}$)
and their masses $M_1$ and $M_2$ (or equivalently $M_1$ and
$r\equiv M_2/M_1$ ).
As the heavier inflaton oscillates, $\mid \dot\phi_2
\mid \propto a^{-\frac{3}{2}}$, $V(\phi_2)\propto a^{-3}$, and
becomes negligible, one has
$\dot\sigma=\dot\phi_1$ and $V(\sigma)= V(\phi_1)$. Therefore, the
value of $\sigma$ can be set equal to $\phi_1$ and they would have
the same potentials. In Fig.{\ref{vfai} we show the effective
potential $V(\sigma)$ as well as $V(\phi_1)$. $V(\sigma)$ becomes
sharper as $r$ increases and the initial value of $\phi_1$ would
also change the shape of the effective potential.

\begin{figure}
\includegraphics[scale=0.6]{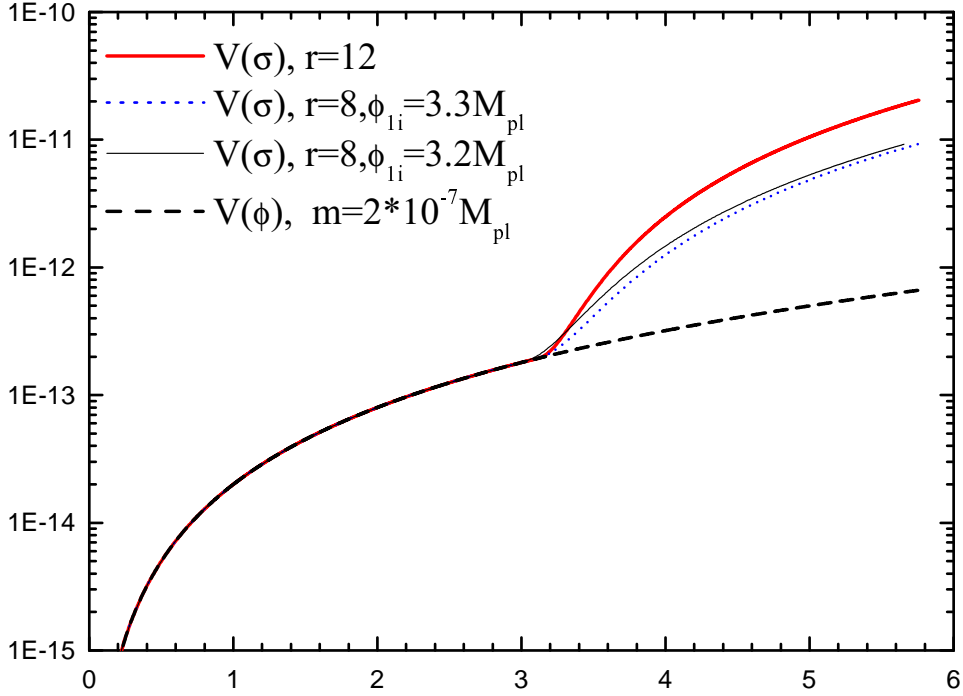}
\caption{\label{vfai} Effective potentials $V(\sigma)$ together
with $V(\phi_1)$. The horizontal axis is the value of inflaton
$\phi_1$ or $\sigma$, in unit of $M_{pl}$. The vertical axis
delineates the inflaton potential, in unit of $M^4_{pl}$. }
\end{figure}

We notice, from Fig. \ref{vfai}, $\dot{\sigma}$ achieves a large
value during the transition time and the scalar power perturbation is
suppressed via the slow-rolling(SR) formula $P_S \propto
(\frac{H^2}{2 \pi \dot{\sigma}})^2$. The SR parameters $\epsilon$
and $\delta$ during the transition are
\begin{equation}
\label{eq:SR1}
 \epsilon \equiv -\frac{\dot{H}}{H^2}= 4\pi G (\frac{\dot{\sigma}}{H})^2
 \approx \frac{3}{2}\frac{\dot{\phi_2}^2}{\rho_{\phi_1}+\rho_{\phi_2}}
 ~,
\end{equation}
and
\begin{equation}
\label{eq:SR2}
 \delta \equiv \frac{\ddot{\sigma}}{H\dot{\sigma}}
 = \frac{\dot{\phi_1}\ddot{\phi_1}+
 \dot{\phi_2}\ddot{\phi_2}}{H(\dot{\phi_1}^2+\dot{\phi_2}^2)}
 \approx  -\frac{3\dot{\phi_2}^2}{\dot{\phi_1}^2+\dot{\phi_2}^2}
 ~.
\end{equation}
We notice that when $\phi_2$ oscillates, $\rho_{\phi_2}\sim
\dot{\phi_2}^2 \propto a^{-3}$, $\epsilon$ and $-\delta$ reach
their local maximum values. One can also
find the maximum value $(-\delta)_{max}>\epsilon_{max}$. In the
extreme limit when $V(\phi_1)$ is negligible during the transition
one has $(-\delta)_{max}=3$ and $\epsilon_{max}=1.5$. Regarding
the fore-mentioned four parameters, the ratio $r\equiv M_2/M_1$ and the
initial value of $\phi_1$, determine the locations and values of
$-\delta_{max}$ and $\epsilon_{max}$. The maximal values are mainly
determined by $r$. If the ratio $r$ is too small (e.g.
$1\leq r \lesssim 3$) the above picture cannot be realized
because both fields would take effect during inflation and neither
is negligible. While $r$ is too large (e.g. $ r\gtrsim
11$) one gets $1+\epsilon+\delta< 0$ during the transition and
superhorizon effects\cite{Leach:2000yw} would take place. The
perturbations do get suppressed at some smaller $k$ but enhanced
around certain larger $k$ values. Under such circumstances the
whole effect might be negative to achieve small CMB TT quadrupole.
The need that $P_S(k)$ be suppressed at small $k$ requires some
tuning of the initial value of $\phi_1$. $M_1$ determines the amplitude of the
perturbation and is normalized by the current observations. The
initial value of $\phi_2$ is arbitrary with a weak prior to provide
enough number of $e-folding$ to solve the flatness problem.

As our model parameters lie in the region 
where SR approximation does not work well, we
calculate the primordial scalar and tensor spectra using mode by mode
integrations\cite{wenbin,wangxl,fengbo2}. We denote the scale
where $P_S$ arrives around its local maximum as $k_f$ and tune
the initial $\phi_1$ to get $N(k_f)\sim 55$.
\begin{figure}
\includegraphics[scale=0.6]{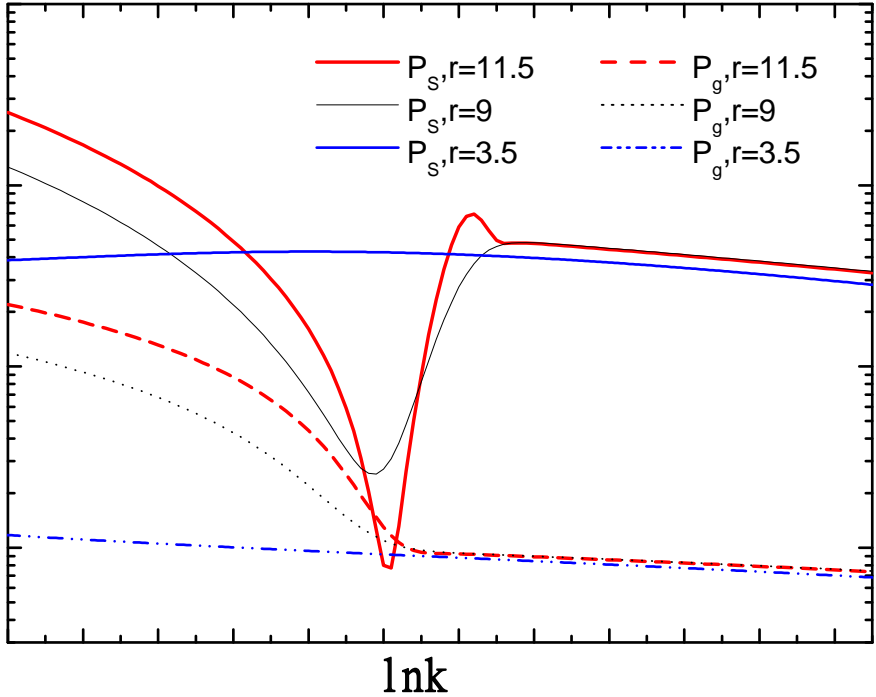}
\caption{\label{PSPG} Primordial scalar $P_s$ and tensor spectra
$P_g$ for $r=3.5$, 9 and 11.5. The overall amplitude can be
normalized by WMAP.}
\end{figure}
In Fig. \ref{PSPG} we show the numerical results of the scalar and
tensor spectra for $r=3.5$, 9 and 11.5. One can see that, for $r=3.5$,
the spectra is almost featureless while  well suppressed scalar
spectra have been generated for $r=9$ and 11.5. For the example of
$r=11.5$ $P_S$ is enhanced around $k_f$ due to the superhorizon
contributions\cite{Leach:2000yw}.

 We then fit the resulting primordial spectra to the
current WMAP TT and TE data. As shown in Refs.\cite{Lyth
reports,fgw}, in such inflation models one cannot know the exact
values of $k_f$ due to the uncertainty in the details of reheating.
So $\ln k_f$
is another parameter in our model. Our fitting is similar to
Ref.\cite{fengbo2}:  We fix  $\Omega_bh^2=0.022$, $\Omega_{m}h^2
= 0.135$, $\tau_c=0.17$, $\Omega_{\rm tot}=1$ \cite{Spergel}
and set $\Omega_{\Lambda}$ and $\ln k_f$ as free parameters in our
fit. Denoting $k_c=7.0\times 70./3/10^5\approx 1.6\times 10^{-3}$
Mpc$^{-1} $, we vary grid points with ranges $[0.68,0.77]$, and
$[-3,5.]$ for $\Omega_{\Lambda}$ and $\ln (k_f/k_c)$, respectively.
$M_2/M_1$ varies from 3.5 to 12 in step of 0.5.
At each point in the grid we use subroutines derived from those
made available by the WMAP team to evaluate the likelihood
with respect to the WMAP TT and TE data \cite{Verde}. The overall
amplitude of the primordial perturbations has been used as a
continuous parameter.

\begin{figure}
\includegraphics[scale=0.6]{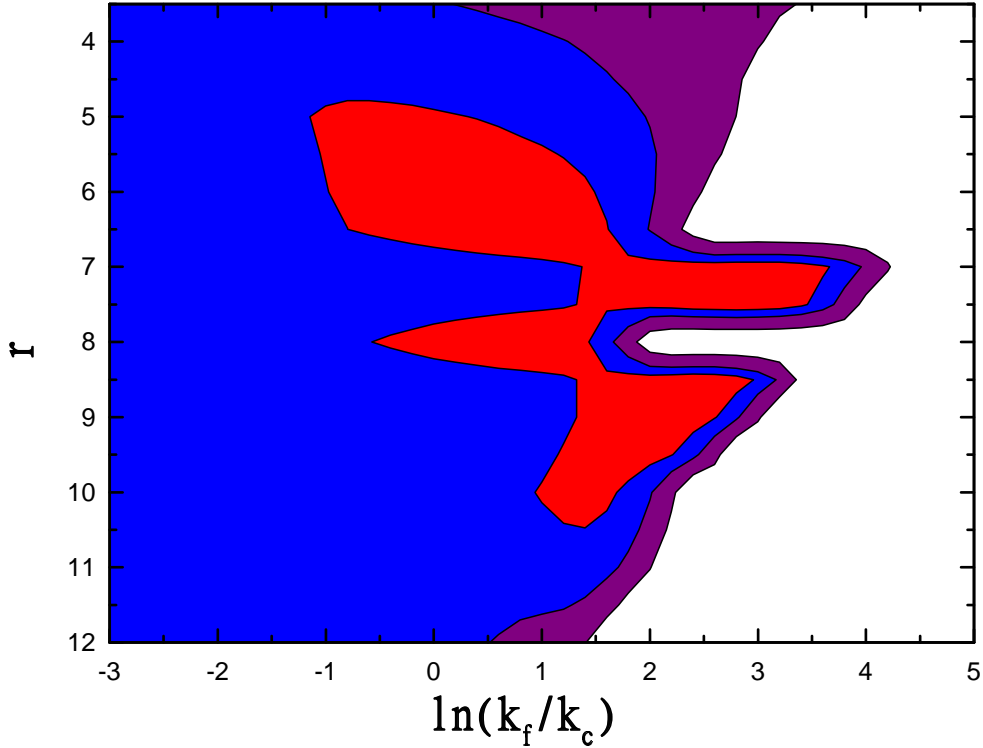}
\caption{\label{cont}   Two-dimensional contours in the r--
$\ln(k_f/k_c)$ plane for our grids of model. $k_c \approx 1.6
\times 10^{-3}$ Mpc$^{-1} $. The regions of different color show
$1.1\sigma$, 2 and $3\sigma$ confidence respectively. }
\end{figure}

In Fig.~\ref{cont} we plot the resulting $\chi^2$ values as
functions of $r$ and $\ln(k_f/k_c)$. The contours shown are for
$\Delta\chi^2$ values giving 1.1, 2, and 3 $\sigma$ contours for
two parameter Gaussian distributions. As the location is rather
hard to be fixed at exactly $N(k_f)=55$, the figure is not very smooth as
expected.
\begin{figure}
\includegraphics[scale=0.6]{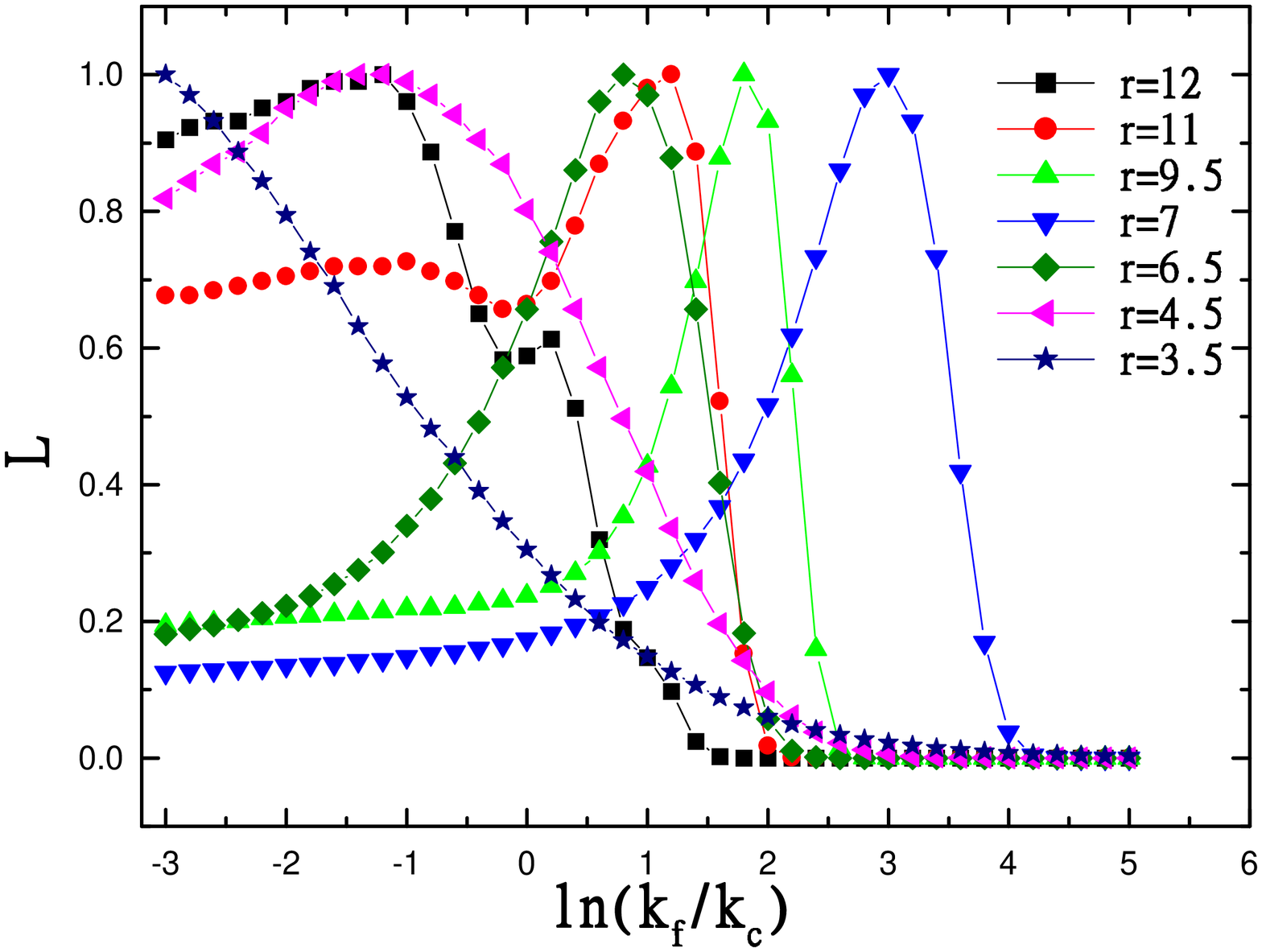}
\caption{\label{L0cont}   One-dimensional marginalized
distributions of $\ln(k_f/k_c)$ for $r=3.5$, 4.5, 6.5, 7, 9.5, 11
and 12.  }
\end{figure}
Our main intention is to see how the primordial spectrum with a
feature is favored by WMAP. This can be also seen in the
one-dimensional marginalized distribution of $\ln(k_f/k_c)$ for
each $r$. To see clearly how the feature is favored, we do
not marginalize over $r$ and show some of them in
Fig.~\ref{L0cont}. For $r=3.5$, $k_f\sim 0$ is favored  and when
$r=7$, $\ln(k_f/k_c)=3$ is favored at around $2\sigma$. $k_f\sim
0$ is excluded at less than $1\sigma$ for $r=4.5$ where $P_S$ is
not suppressed enough around $k_f$. While for $r\gtrsim 11$, $P_S$
is enhanced around $k_f$ and although nonzero $k_f$ is favored for
shown examples, $k_f\sim 0$ is excluded at less than $1\sigma$. We
find that, for $6\lesssim r \lesssim 10$, nonzero $k_f$ is
favored at $\gtrsim 1.5\sigma$. For the investigated parameter
space with $3.5\lesssim r \lesssim 12$ we
have $P_S(0.05/Mpc)=2.46\sim 2.59 \times 10^{-9}$ at $2\sigma$
level. This gives $M_1\sim 1.7 \times 10^{13}$ GeV.

A detailed analysis gives
the e-folds number $N(k)$ before the end of inflation\cite{Lyth reports,fgw}:
 \begin{eqnarray}
N(k)=
 60.56-\ln h-\ln\frac{k}{a_0H_0}-\ln\frac{10^{16}GeV}{\rho(k)^{1/4}}
+\ln\frac{\rho(k)^{1/4}}{\rho_{end}^{1/4}}-
\frac{1}{3}\ln\frac{\rho_{end}^{1/4}}{\rho_{RH}^{1/4}}\ \ , \end{eqnarray}
where $\rho(k)$, $\rho_{end}$ denote  the inflaton potential at $k=aH$
and at the end of inflation respectively, $\rho_{RH}$ is the energy density
when reheating ends, resuming a standard big bang
evolution. Since in our case there is a preferred scale $\ln
 (k_f/k_c)$ while $N(k_f)$ is fixed around 55, the reheating energy may
 be determined by the current observations. However, one can see that the
 location of $k_f$ is mainly determined by the initial value of $\phi_1$. Once
the initial $\phi_1$ changes, $N(k_f)$ will change and the resulting
 $\rho_{RH}$ would be different. We show the case in Fig.
 \ref{1dboth} as an example. For $r=8$, $\phi_{1i}=3.2$ and $3.3M_{pl}$
 lead to $N(k_f)=54.34$ and 59.06 respectively.
We get $P_S  \sim 2.5 \times 10^{-9}$  at 0.05 Mpc$^{-1}$, the
resulting $\ln (k_f/k_c)=(-1.7,1.3)$ and $(0.1,1.9)$ at $2\sigma$
respectively. We also have $h\approx 0.73$,
$\rho^\frac{1}{4}(0.05/Mpc)\sim 1.8\times 10^{-3}M_{pl}$ and
$\rho^\frac{1}{4}_{end}\sim 5.1\times 10^{-4}M_{pl}$. Taking these
to the models we get $\rho^\frac{1}{4}_{RH}=(8.5\times 10^4 GeV,
6.9\times 10^{8} GeV)$ and $(2.6\times 10^{13} GeV, 5.8\times
10^{15} GeV)$ at $2\sigma$ for the two different $\phi_{1i}$.
Therefore, the reheating temperature is fully
correlated with initial $\phi_1$ in this model.

\begin{figure}
\includegraphics[scale=0.6]{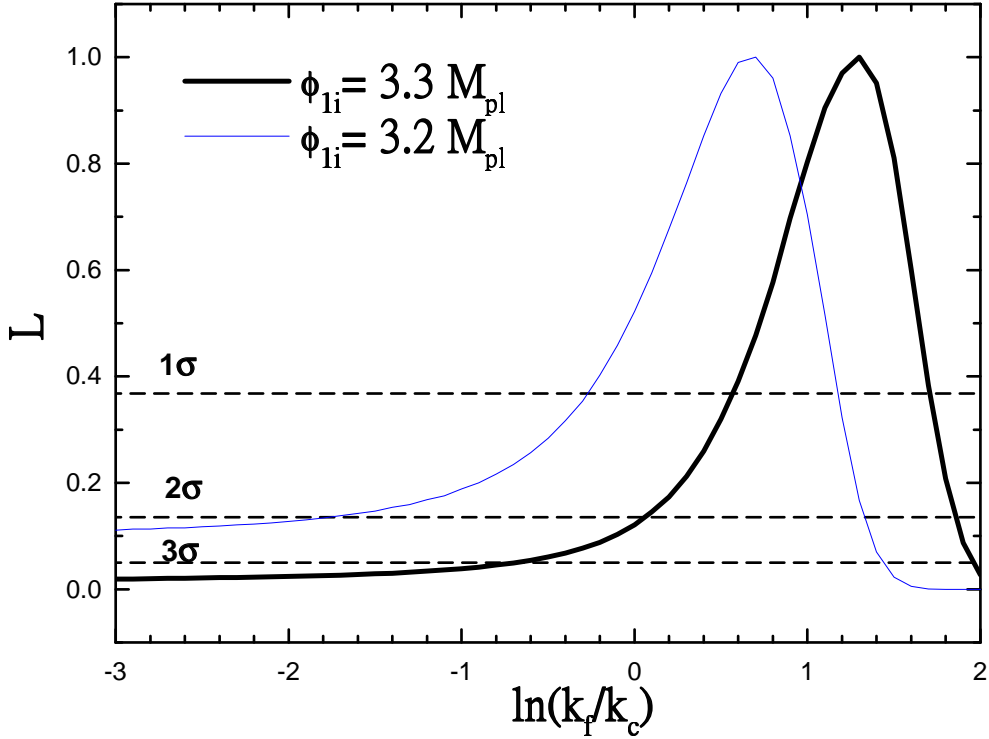}
\caption{\label{1dboth} One dimensional likelihoods of r=8,
$\phi_{1i}=3.2$ and 3.3 $M_{pl}$. }
\end{figure}

We get our minimum $\chi^2=1429.1$ when $r=8.5$ and $\ln
(k_f/k_c)=2.4$. When compared with the standard power-law
$\Lambda$CDM model, we have minimum $\chi^2=1432.7$ and $\Delta
\chi^2 = -3.6$. For the single field chaotic inflation we get
minimum $\chi^2=1432.9$, with $\Delta \chi^2 = -3.8$. However, in
the sneutrino inflation, we have to set
$\rho_{RH}^\frac{1}{4}\lesssim 10^{10}$ GeV due to the
gravitino problem\cite{gravitino}. In this case, we get
$N(k_c)\lesssim 55.5$ and minimum $\chi^2=1433.2$, which gives
$\Delta \chi^2 = -4.1$. In addition, there are only two
parameters, the mass and $\ln (k_f/k_c)$, in the single field
sneutrino inflation model. This indicates our double sneutrino
inflation is favored at $\sim 1.5 \sigma$ by WMAP than the single
field sneutrino inflation. In Fig. \ref{Cl} we show the resulting
CMB TT multipoles and two-point temperature correlation function
for single and double field sneutrino inflation in our parameter
space. One can see that the resulting CMB TT quadrupole and the
correlation function at $\theta\gtrsim 60^\circ $ are much better
suppressed in the double sneutrino inflation than in the single
sneutrino model. In fact, the spectrum of the single field
sneutrino inflation is equivalent to that in our double case with
$r=1$ and $\phi_{1i}=\phi_{2i}$. In this sense we get $6\lesssim
r\lesssim 10$ is favored at $\gtrsim 1.5 \sigma$ ($\Delta \chi^2
\lesssim -2.3$) than $r=1$ in double sneutrino inflation.
\begin{figure}
\includegraphics[scale=0.6]{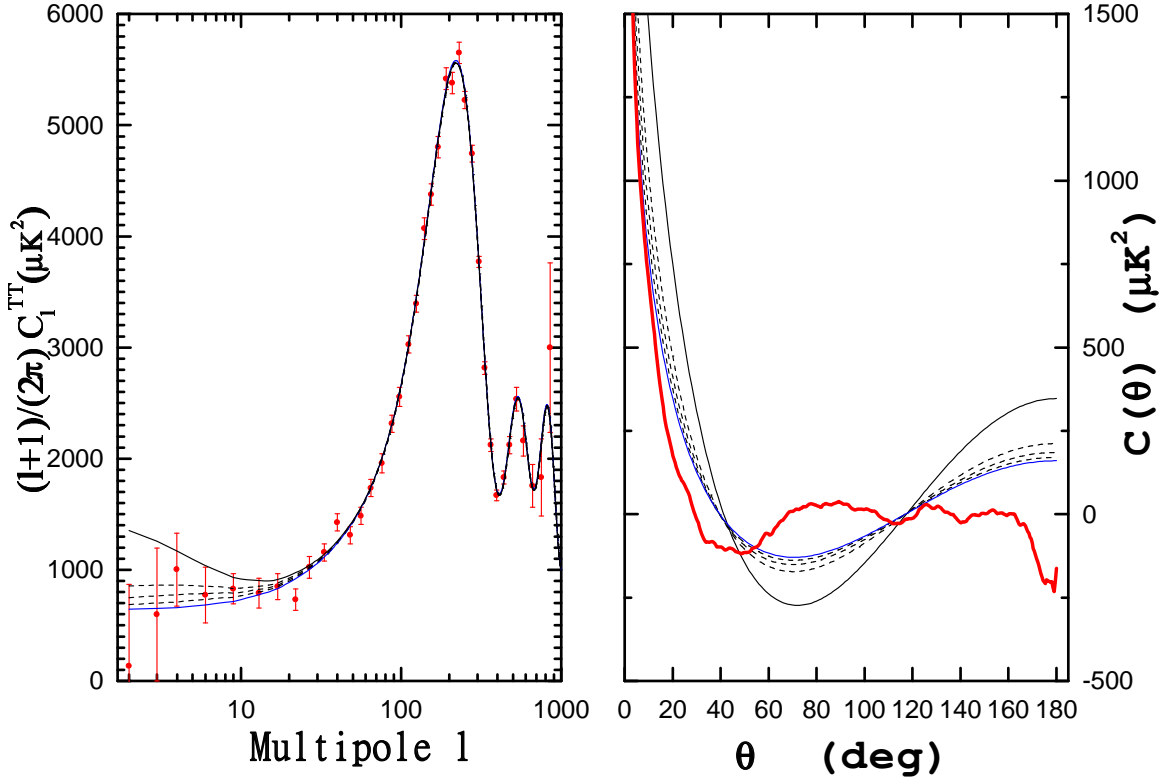}
\caption{\label{Cl} CMB anisotropy and two-point temperature
correlation function for single and double field sneutrino
inflation. Left: From left top to bottom, the lines stand for
single sneutrino inflation, double sneutrino inflation with $\ln
(k_f/k_c)=3.0$, 3.2, 3.4 and 3.6. $r$ is fixed at 8.5. Right: From
right top to bottom, the lines stand for single sneutrino
inflation, double sneutrino inflation with $\ln (k_f/k_c)=3.0$,
3.2, 3.4 and 3.6 and the WMAP released data.}
\end{figure}

Finally, it is worth mentioning that we have also considered
a double inflaton model with quartic potential\footnote{The
quartic term of sneutrino is absent in the minimal supersymmetric
seesaw mechanism. These terms can arise if the RH neutrino Majorana
mass is produced in the superpotential $\lambda \Phi N N$, with
$\Phi$ another superfield whose vacuum expectation value generates
the Majorana mass.}
\begin{equation}\label{potentlf4} V(\phi_1,\phi_2)=
 \lambda_1\phi_1^4 +  \lambda_2\phi_2^4~.
\end{equation}
As we known, the quartic potential $\lambda \phi^4$
is disfavored by the current WMAP and LSS observations,
because it has a larger tensor perturbation.
Peiris {\it et al.} \cite{Peiris} fix the number of e-folding at
50 and find $\lambda \phi^4$ inflation model is excluded at more
than $3\sigma$ by WMAP and 2DFGRS data. WMAP alone excludes
$\lambda \phi^4$ inflation at more than $99\%$ confidence level
when $N\sim 50$. The discrepancy between the theoretical predictions
and observations comes mainly from the contributions of small CMB
multipoles. In the double inflaton quartic model,
the CMB quadruples can also be well suppressed and the model is
also favored by WMAP. We fix $N(k_f)=50$ and run two codes,
one with $\lambda_2/\lambda_1=6400$ and the other with
$\lambda_2/\lambda_1=3600$ and fit the primordial scalar and
tensor spectra to WMAP TT and TE data. We get minimum
$\chi^2=1427.9$ and $1428$ respectively. They work better than
the double quadratic sneutrino inflation. Reheating temperature
in this case
cannot be restricted from WMAP, as shown in Ref.\cite{liddle}.

\section{phenomenology}

In the minimal
seesaw mechanism, the right-handed sector is least known.
However, in the double sneutrino inflation model,
two neutrino masses $M_1$ and $M_2$ are constrained by the WMAP
as shown in the previous section. In the following,
we will study the phenomenological implications of this model,
including the reheating temperature, leptogenesis, lepton flavor
violation and neutrinoless double beta decay.

\subsection{Parameterization of the minimal seesaw model}

In this subsection we present our convention and
parameterization of the minimal supersymmetric seesaw model.
At the energy scales above the RH neutrino masses, the
superpotential of the lepton sector is given by
\begin{equation}
\label{lag}
W=Y_L^{ij*}\hat{H}_1\hat{L}_i\hat{E}_j+Y_N^{ij*}\hat{H}_2\hat{L}_i\hat{N}_j
+\frac{1}{2}M_R^{ij*}\hat{N}_i\hat{N}_j+\mu \hat{H}_1\hat{H}_2\ \ ,
\end{equation}
where $Y_L$ and $Y_N$ are the charged lepton and
neutrino Yukawa coupling matrices, respectively,
$M_R$ is the Majorana mass matrix for the right-handed neutrinos,
with $i$ and $j$ being the generation indices.

Generally, $Y_L$ and $Y_N$ can not be diagonalized simultaneously.
This mismatch leads to the lepton flavor violating (LFV) interactions.
The three matrices $Y_L$, $Y_N$ and $M_R$ can be diagonalized by
\begin{eqnarray}
\label{ul}
Y^\delta_L&=&U_L^\dagger Y_L U_R\ \ ,\\
\label{vl}
 Y^\delta_N&=&V_L^\dagger Y_N V_R\ \ ,\\
\label{x1}
 M_R^\delta&=&X V_R^T M_R V_R X^T\ \ ,
\end{eqnarray}
respectively, where $U_{L,R}$\ , $V_{L,R}$ and $X$ are all unitary
matrices.

We can define the lepton flavor mixing matrix $V$,
the analog to the Kobayashi-Maskawa matrix $V_{KM}$ in the quark
sector, as
\begin{equation}
\label{vd} V=U_L^\dagger V_L\ .
\end{equation}
$V$ is determined by the left-handed mixing of the Yukawa
coupling matrices $Y_L$ and $Y_N$, and only exists above
the energy scales $M_R$. We will see below that this matrix determines
the LFV effects in the supersymmetric seesaw model at low energies.

We then rotate the bases of $\hat{L}$, $\hat{E}$ and $\hat{N}$ to make both
$Y_L$ and $M_R$ diagonal. On this basis, $Y_N$ can be written in a
general form as
\begin{equation}
\label{ydirac}
Y_N=VY^\delta X^T\ .
\end{equation}
By adjusting the phases of the superfields, $V$ is a CKM-like mixing
matrix with one physical CP phase, and $X$ has the form
\begin{equation}
X=\left(\begin{array} {ccc} 1 & & \\ & e^{i\alpha} & \\
&& e^{i\beta} \end{array} \right) \tilde{X}
\left(\begin{array} {ccc} 1 & & \\ & e^{i\rho} & \\
&& e^{i\omega} \end{array} \right) \ ,
\end{equation}
where $\alpha$, $\beta$, $\rho$ and $\omega$ are Majorana
phases and $\tilde{X}$ is a CKM-like mixing matrix with another
Dirac CP phase.
It is then easy to count that there are 18 parameters to
parametrize the minimal seesaw mechanism, which include
6 Yukawa coupling constants (or mass) eigenvalues in $Y_N$ and $M_R$,
6 mixing angles and 6 CP phases in $V$ and $X$.

At low energies, the heavy RH neutrinos are integrated out and
the Majorana mass matrix for the left-handed neutrinos is given by
\begin{equation}
\label{seesaw}
m_\nu=-m_N\frac{1}{M_R}m_N^T\ ,
\end{equation}
where $m_N=Y_N v\sin\beta$ is the neutrino Dirac mass matrix,
with $v$ being the vacuum expectation value (VEV) of the Higgs boson.
$m_\nu$ can be diagonalized by
\begin{equation}
U_\nu^\dagger m_\nu U_\nu^*=m_{\nu}^\delta\ ,
\end{equation}
where $U_\nu=\tilde{U}_\nu \cdot\text{diag}(1,e^{i\eta},e^{i\xi})$
is the MNS mixing matrix\cite{MNS}, with $\eta$, $\xi$ being
low energy Majorana CP phases.
$U_\nu$ describes the neutrino mixing at low energies,
which is different from the high energy mixing
matrix $V$ defined in Eq. (\ref{vd}).
From Eq. (\ref{seesaw}) we can see that
$m_\nu$ is related to  all the 18 seesaw parameters.
However, measuring $m_\nu$ at low energy only determines 9 of
the 18 seesaw parameters. We will see below that leptogenesis and
lepton flavor violation are related to different combinations
of the 18 seesaw parameters and can provide different information
to determine the seesaw parameters from the
$\nu$-oscillation and LFV observations.

We can rewrite the seesaw formula Eq. (\ref{seesaw}) in another form
\begin{equation}
U_\nu \sqrt{m_\nu^\delta} \left(U_\nu \sqrt{m_\nu^\delta}\right)^T=
-Vm_N^\delta X^T\frac{1}{\sqrt{M_R^\delta}}
\left(Vm_N^\delta X^T\frac{1}{\sqrt{M_R^\delta}}\right)^T\ ,
\end{equation}
from which $m_N$ can be solved in terms of the left-
and right-handed neutrino masses,
\begin{equation}
\label{mn}
m_N^\delta =V'^\dagger \sqrt{m_\nu^\delta} \hat{O}^T\sqrt{M_R^\delta}
\tilde{X}^*\ ,
\end{equation}
where $\hat{O}$ is an arbitrary orthogonal $3\times 3$ matrix\cite{casas}
and $V'=\tilde{U}_\nu^\dagger V$. In the above equation we have
absorbed all the 6 Majorana CP phases in the diagonal eigenvalue
matrices: two low energy Majorana phases, $\eta$, $\xi$, are absorbed by
$\sqrt{m_\nu^\delta}$ and the four high energy Majorana phases, $\alpha$,
$\beta$, $\rho$, $\omega$,
are absorbed by $m_N^\delta$ and $\sqrt{M_R^\delta}$.
We will use this equation repeatedly in the following discussions.

\subsection{The reheating temperature}

The lightest sneutrino $\tilde{N}_1$ begins to oscillate
when the Hubble expansion rate $H\sim M_1$ and decays
at $H\sim \Gamma_{\tilde{N}_1}$.
The Universe is then reheated by the relativistic decay products.
The reheating temperature is approximately determined by
\begin{equation}
T_{RH}\approx \left( \frac{90}{\pi^2 g_*} \right)^\frac{1}{4}
\sqrt{\Gamma_{\tilde{N}_1} M_P}\ ,
\end{equation}
where $g_*$ is the number of the effective relativistic degrees
of freedom in the reheated Universe, $M_P=1/\sqrt{8\pi G_N}
\simeq 2.4\times 10^{18} GeV$ is the Planck scale, and
\begin{equation}
\label{gamman1}
\Gamma_{\tilde{N}_1}=\frac{1}{4\pi}(Y_N^\dagger Y_N)_{11}
M_1\ ,
\end{equation}
is the width of the lightest sneutrino $\tilde{N}_1$, if it couples
to other matter only through the Yukawa coupling in Eq. (\ref{lag}).
Taking $M_1\approx 1.7\times 10^{13} GeV$ and
$T_{RH}\sim 10^{10} GeV$, we get $(Y_N^\dagger Y_N)_{11}$
should be as small as $\mathcal{O}(10^{-10})$.

The reheating temperature (as well as leptogenesis) is related to
the RH mixing of $Y_N$ and put strong constraints on this mixing
matrix. Using Eq. (\ref{ydirac}), we have
\begin{equation}
(Y_N^\dagger Y_N)_{11}=(X^* ({Y^\delta})^2 X^T)_{11}
=(\tilde{X}^* ({Y^\delta})^2 \tilde{X}^T)_{11}
=|\tilde{X}_{1i}|^2 Y_i^2 \ .
\end{equation}
The elements $|\tilde{X}_{1i}|$ can be parametrized by two
mixing angles, $\theta_{1,2}$. We then get
\begin{equation}
\label{ydy11}
(Y_N^\dagger Y_N)_{11}= c_1^2c_2^2Y_1^2+c_1^2s_2^2Y_2^2+s_1^2Y_3^2\approx
Y_1^2+s_2^2Y_2^2+s_1^2Y_3^2\approx 10^{-10}\ ,
\end{equation}
with $c_{i}=\cos{\theta_i}$, $s_{i}=\sin{\theta_i}$.
In the later discussion we will see that $Y_2$ is
$\mathcal{O}(0.1)$ and $Y_3$ is $\mathcal{O} (1)$.
Then we have
\begin{equation}
Y_1^2\lesssim 10^{-10}\ ,\ s_2^2\lesssim 10^{-8}\ ,\
s_1^2\lesssim 10^{-10}\ .
\end{equation}

Since $\theta_{1,2}$ are extremely small,
$\tilde{X}$ can be given in a quite simple form as
\begin{equation}
\label{xw}
\tilde{X}\approx \left(
\begin{array}{ccc} 1 & s_2 & \hat{s}_1 \\
-(c_3s_2+s_3\hat{s}_1^*) & c_3 & s_3 \\
s_3s_2-c_3\hat{s}_1^* & -s_3 & c_3 \end{array} \right)\ ,
\end{equation}
where $\hat{s}_1=s_1 e^{i\delta}$.
%We will use this form of $\tilde{X}$
%when we discuss the leptogenesis.

Using Eqs. (\ref{mn}) and (\ref{gamman1}), we have
\begin{eqnarray}
\label{mnd1}
(m_N^\dagger m_N)_{11}&=&\frac{4\pi \Gamma_{\tilde{N}_1}}{M_1}
(v\sin\beta)^2\approx 4\times 10^{-6} \left( \frac{T_{RH}}{10^{10} GeV}
\right)^2 GeV^2\nonumber \\
& \approx& M_1 m_{\nu_1}|\hat{O}_{11}|^2+
120.7 GeV^2|\hat{O}_{12}|^2+850 GeV^2|\hat{O}_{13}|^2 \ ,
\end{eqnarray}
where we have assumed $\sin\beta \approx 1$ for large $\tan\beta$,
and $m_{\nu_2}\approx \sqrt{\Delta m_{sol}^2}\approx 7.1\times 10^{-3} eV$
and $m_{\nu_3}\approx \sqrt{\Delta m_{atm}^2}\approx 0.05 eV$.
From the above equation we can see that $\hat{O}_{12}$ and $\hat{O}_{13}$
have to be negligibly small. We will set these two elements zero
and write $\hat{O}$ as
\begin{equation}
\hat{O}=\left( \begin{array}{ccc} \pm 1 & & \\  & \hat{c} & \hat{s} \\
&-\hat{s}&\hat{c} \end{array} \right)\ ,
\end{equation}
where $\hat{c}=\cos\theta_T$, $\hat{s}=\sin\theta_T$ with
$\theta_T$ being an arbitrary complex angle.
(It should be noted that $\hat{O}_{12}$ and $\hat{O}_{13}$
can not be exactly zero, since if they are zero the
first-generation right-handed
(s)neutrino decouples from the other two generations and no
lepton number asymmetry can be induced when it decays. However,
the tiny mixing has no effect on lepton flavor violation
and we can ignore them safely
when discussing LFV.)

From Eq. (\ref{mnd1}) we can estimate that
\begin{equation}
\label{mnu1}
m_{\nu_1}\approx
\left\{\begin{array}{l}
2.\times 10^{-10} eV,\ \ \text{for}\ \ T_{RH}=10^{10} GeV\ , \\
2.\times 10^{-12} eV,\ \ \text{for}\ \ T_{RH}=10^{9} GeV\ .
\end{array}\right.
\end{equation}
This estimation is correct when the last two terms are much
smaller than the first  one in the second line of Eq. (\ref{mnd1}),
or, equivalently, the $Y_1$ term dominants the others in Eq.
(\ref{ydy11}). In the following discussion for leptogenesis we
will see that this is a quite natural situation.

\subsection{Leptogenesis}

Since the reheating temperature, $T_{RH}$, is far below
the lightest RH (s)neutrino mass, $M_1$, leptogenesis
arises dominantly from direct cold sneutrino decays,
with negligible thermal wash-out effects.
In this case, the baryon asymmetry is given by\cite{hamaguchi}
\begin{equation}
Y_B\equiv \frac{n_B}{s}=a \frac{3}{4}\epsilon_1\frac{T_{RH}}{M_1}\ ,
\end{equation}
where $a=-8/23$ is the ratio of baryon to lepton asymmetry
balanced by the ``sphaleron'' process.
In order to produce the observed baryon asymmetry in the Universe,
$Y_B\sim  10^{-10}$, we require the sneutrino decay
asymmetry $\epsilon_1\sim -10^{-6}$.
The asymmetry $\epsilon_1$ is given by
\begin{equation}
\epsilon_1 \approx -\frac{3}{16\pi}\frac{1}{(Y_N^\dagger Y_N)_{11}}
\sum_{i=2,3} \text{Im}\left[ (Y_N^\dagger Y_N)_{1i} \right]^2\frac{M_1}{M_i} \ .
\end{equation}
Using the expression for $\tilde{X}$ in Eq. (\ref{xw}) and
the large hierarchy between $Y_1$ and $Y_2$, $Y_3$
we get
\begin{eqnarray}
(Y_N^\dagger Y_N)_{12}&\approx& (s_2Y_2^2+\hat{s}_1^*s_3Y_3^2)e^{i\alpha}\ ,\nonumber\\
(Y_N^\dagger Y_N)_{13}&\approx& (-s_2s_3Y_2^2+\hat{s}_1^*Y_3^2)e^{i\beta}\ .
\end{eqnarray}

We will discuss two simple cases to illustrate some quantitative
features of the seesaw parameters required by leptogenesis. We will
see that, in Eq. (\ref{ydy11}), the $Y_2$ and $Y_3$ terms should be smaller
than the $Y_1$ term in order to produce the lepton number asymmetry at
the correct order.
\begin{itemize}
\item{Case I, $s_1Y_3^2 \ll s_2Y_2^2$}

In this case the expression for $\epsilon_1$ is simplified as
\begin{eqnarray}
\epsilon_1 &\approx & -\frac{3}{16\pi}\frac{1}{(Y_N^\dagger Y_N)_{11}}
 \left[ ({s_2Y_2^2})^2\sin2\alpha \frac{M_1}{M_2} +
(s_2s_3Y_2^2)^2\sin2\beta\frac{M_1}{M_3}\right]\nonumber \\
&\sim & -10^{-4} \cdot \frac{s_2^2Y_2^2}{Y_1^2+s_2^2Y_2^2}\cdot\sin2\alpha \ .
\end{eqnarray}
When deriving the second line we have assumed that $\alpha$ and $\beta$ are
of the same order and $\frac{M_1}{M_3}\ll \frac{M_1}{M_2}\sim Y_2\sim 0.1$
and $s^2_1Y_3^2 \ll s^2_2Y_2^2$.
If the CP phases are of order 1,
$s_2^2Y_2^2/Y_1^2$ should be at the order of about $10^{-2}$.
Actually, this case corresponds to the maximal asymmetry given
by $|\epsilon_1^{max}|\approx \frac{3}{16\pi}\frac{M_1\sqrt{\Delta
m_{sol}^2}}{v^2}\sim 10^{-4}$ \cite{buch}. In this case, the CP phase
or, $s_2^2Y_2^2/(Y_N^\dagger Y_N)_{11}$,
has to be at the order of $\mathcal{O}(10^{-2})$.

\item{Case II, $s_1\sim s_2$}

In this case we can simplify the expression for $\epsilon_1$ as
\begin{eqnarray}
\epsilon_1 &\approx & -\frac{3}{16\pi}\frac{1}{(Y_N^\dagger Y_N)_{11}}
 \left[ ({s_1s_3Y_3^2})^2\sin2\alpha' \frac{M_1}{M_2} +
(s_1Y_3^2)^2\sin2\beta'\frac{M_1}{M_3}\right]\nonumber \\
&\sim & -10^{-3} \cdot \frac{s_1^2Y_3^2}{Y_1^2+s_1^2Y_3^2}\cdot
(\sin2\alpha'+\sin2\beta') \ ,
\end{eqnarray}
where we have used the fact that $s_3 \sim \sqrt{M_2/M_3} $ if
$\theta_T$ is of order $1$, $s^2_1Y_3^2 \gg s^2_2Y_2^2$
and $\alpha'=\alpha-\delta$,
$\beta'=\beta-\delta$. Similar to Case I,
we get that $s_1^2Y_3^2/Y_1^2$ should be at the order of
$10^{-3}$ if the CP phases are
of order 1. In this case, the maximal asymmetry is
$|\epsilon_1^{max}|\approx \frac{3}{16\pi}\frac{M_1\sqrt{\Delta
m_{atm}^2}}{v^2}\sim 10^{-3}$.
\end{itemize}

Certainly, it is possible that the contributions to
$(Y_N^\dagger Y_N)_{11}$ from $Y_2$ and
$Y_3$ in Eq. (\ref{ydy11}) are of the same order.
In this case we also expect that these values
be correct as an estimate of the order of magnitude, i.e.,
$s^2_1Y_3^2 \sim s^2_2Y_2^2 \ll Y^2_1$.
This analysis justifies our guess in the last
subsection that the $Y_1$ term gives the dominant contribution
in the process of reheating the Universe.
Conversely, if the $Y_2$ or $Y_3$ term gives dominant
contribution, the CP phases
have to be fine tuned to the order of $10^{-2}$
and $10^{-3}$ respectively, in order not to create too much lepton number
asymmetry and $m_{\nu_1}$ in Eq. (\ref{mnu1}) will be even smaller.

\subsection{Lepton flavor violation and
muon anomalous magnetic moment}

We have shown that leptogenesis is associated with the high
energy mixing angles and CP phases in the unitary matrix $X$.
Generally, leptogenesis has no direct relation with the low energy
neutrino phenomena. However, another interesting
phenomena --- the charged lepton flavor violating decays ---
predicted by this
sneutrino inflaton model, can provide constraints
on the seesaw model's parameter space.
The muon anomalous magnetic moment is also considered to constrain
the SUSY parameters.

In a supersymmetric model,
the present experimental limits on the LFV processes has put
very strong constraints on the soft supersymmetry breaking
parameters, with the strongest constraints coming from the
process $\mu\to e\gamma$
(BR$(\mu\to e\gamma) < 1.2 \times 10^{-11}$\cite{mueg}).
It is a usual practice to assume universal soft SUSY
breaking parameters
$m_0$, $m_{1/2}$ and $A_0$ at the SUSY breaking scale ( We take
it the GUT scale here) to
suppress the LFV effects.
%We expect that under such
%hypotheses the SUSY seesaw model predicts the minimum LFV effects,
%besides those contributios from possible flavor structure
%existing at the GUT scale.
However, since there are LFV interactions in the
seesaw models, the lepton flavor violating
off-diagonal elements of $(m_{\tilde{L}}^2)_{ij}$,
the slepton doublet soft mass matrix,
and $({A_e})_{ij}$, the lepton soft trilinear couplings,
can be induced when running the renormalization
group equations (RGEs) for $m_{\tilde{L}}^2$ and
${A_e}$ between $M_{GUT}$ and $M_R$.

The off-diagonal elements of
$(m_{\tilde{L}}^2)_{ij}$ and $({A_e})_{ij}$
can be approximately given by
\begin{eqnarray}
\left(\delta m_{\tilde{L}}^2\right)_{ij}
&\approx& \frac{1}{8\pi^2}
(Y_NY_N^\dagger)_{ij} (3+a^2)m_0^2
\log\frac{M_{GUT}}{{M_R}}\ , \\
\label{dm}
\left(\delta A_e\right)_{ij}
&\approx& \frac{1}{8\pi^2}
(Y_NY_N^\dagger)_{ij} a m_0
\log\frac{M_{GUT}}{{M_R}}\ ,
\end{eqnarray}
where $A_0=am_0$ is the universal trilinear coupling at $M_{GUT}$.
Using Eq. (\ref{ydirac}) we have
\begin{eqnarray}
(Y_NY_N^\dagger)_{ij} &=&
(V(Y_N^\delta)^2V^\dagger)_{ij}
\approx V_{i2}V_{j2}^*Y_2^2+V_{i3}V_{j3}^*Y_3^2\ ,\ \
\text{for}\ M_{GUT} > Q > M_3\\
&=& \sum_{k=1,2}(VY_N^\delta X^T)_{ik}
(X^*Y_N^\delta V^\dagger)_{kj} \nonumber\\
\label{ydy23}
&\approx&
\sum_{l,m=2,3}V_{il}V_{jm}^*Y_lY_m(\delta_{lm}-X_{3l}X_{3m}^*)\ ,\ \
\text{for}\ M_3 > Q > M_2\ .
\end{eqnarray}
The numerical result shows that, since the mixing angles
in $X$ are all small, the LFV effects
are only sensitive to the  left-handed mixing matrix $V$,
while leptogenesis only relies on the right-handed mixing matrix $X$.
Thus, there are no direct relation between the two phenomena in
principle.

We have solved the full coupled RGEs numerically from the GUT scale
to $M_Z$ scale.
At the energy scales below $M_2$ we solve the RGEs for MSSM and below
$M_{SUSY}$ the RGEs return to those of the SM.

In principle, only 9 of the 18 seesaw parameters are
determined in our model, i.e., $m_{\nu_i}$, $M_i$ and 3 low
energy neutrino mixing angles. In order to predict the
branching ratio of the LFV decays, we have to explore a 9-dimensional
parameter space of the unknown variables.
However, from our previous discussions,
we know that the relevant seesaw parameters to LFV are
reduced to only 4 in this model,
which can be chosen as 1 complex angle $\theta_T$,
and 2 CP phases. We can explicitly write the `reduced'
seesaw formula for the 2nd and 3rd generations as
\begin{eqnarray}
\label{2ss}
\left(
\begin{array}{cc} m_{N_2} & \\ & m_{N_3} e^{i\omega'} \end{array}
\right) &=&
V'^\dagger
\left(
\begin{array}{cc} \sqrt{m_{\nu_2}} & \\ & \sqrt{m_{\nu_3}}e^{i\phi_1}\end{array}
\right)
\left(
\begin{array}{cc} \hat{c} & -\hat{s} \\ \hat{s} & \hat{c} \end{array}
\right)
\left( \begin{array}{cc} \sqrt{M_2} & \\ & \sqrt{M_3} e^{i\phi_2}\end{array}
\right)
\tilde{X}\nonumber \\
&=& V'^\dagger \left(
\begin{array}{cc} 1 & \\ & e^{i\phi_1}\end{array}
\right)
\mathcal {M}
\left(
\begin{array}{cc} 1 & \\ & e^{i\phi_2}\end{array}
\right)
\tilde{X}\ ,
\end{eqnarray}
where both $V'$ and $\tilde{X}$ are $2\times 2$ real orthogonal matrices,
determined by diagonalizing the matrix $\mathcal {M}$.
Here, we adopt the running values of $m_{\nu_2}$ and $m_{\nu_3}$
at the scale of $10^{14} GeV$\cite{ratz}.
Once $Y_{2,3}$ and $V=U_\nu V'$ are determined, can we calculate
the LFV branching ratios, BR($l_i\to l_j\gamma$).

The relevant parameters to investigate BR($l_i\to l_j\gamma$)
and $\delta a_\mu$
include the mSUGRA parameters: $m_0$, $m_{1/2}$, $A_0$,
$\tan\beta$, $\text{sgn}(\mu)$ and the seesaw parameters:
$\theta_T$ and $\phi_i$.
Since BR($l_i\to l_j\gamma$) and $\delta a_\mu$ nearly scale with
$\tan^2\beta$ and $\tan\beta$ respectively, we take $\tan\beta=10$
as a representative value.
We fix $A_0=0$ through our calculation since it has small influence
on the numerical results.
The Higgsino mass parameter $\mu > 0$ is assumed, motivated by
the $g_\mu-2$ anomaly.
As for the seesaw parameters, we take
$\Delta m_{sol}^2=5\times 10^{-5} eV^2$,
$\Delta m_{atm}^2=2.5\times 10^{-3} eV^2$,
and $\tan^2\theta_{12}=0.42$, $\sin^22\theta_{23}=1$,
$ 0 < \theta_{13} < 0.2$ from the neutrino oscillation
experiments.
We fix $M_1=1.7\times 10^{13} GeV$, $M_2=10^{14} GeV$
and $4\times 10^{14} GeV < M_3 < 1\times 10^{16} GeV$ for
RH heavy Majorana neutrinos.

\begin{figure}
\includegraphics[scale=0.5]{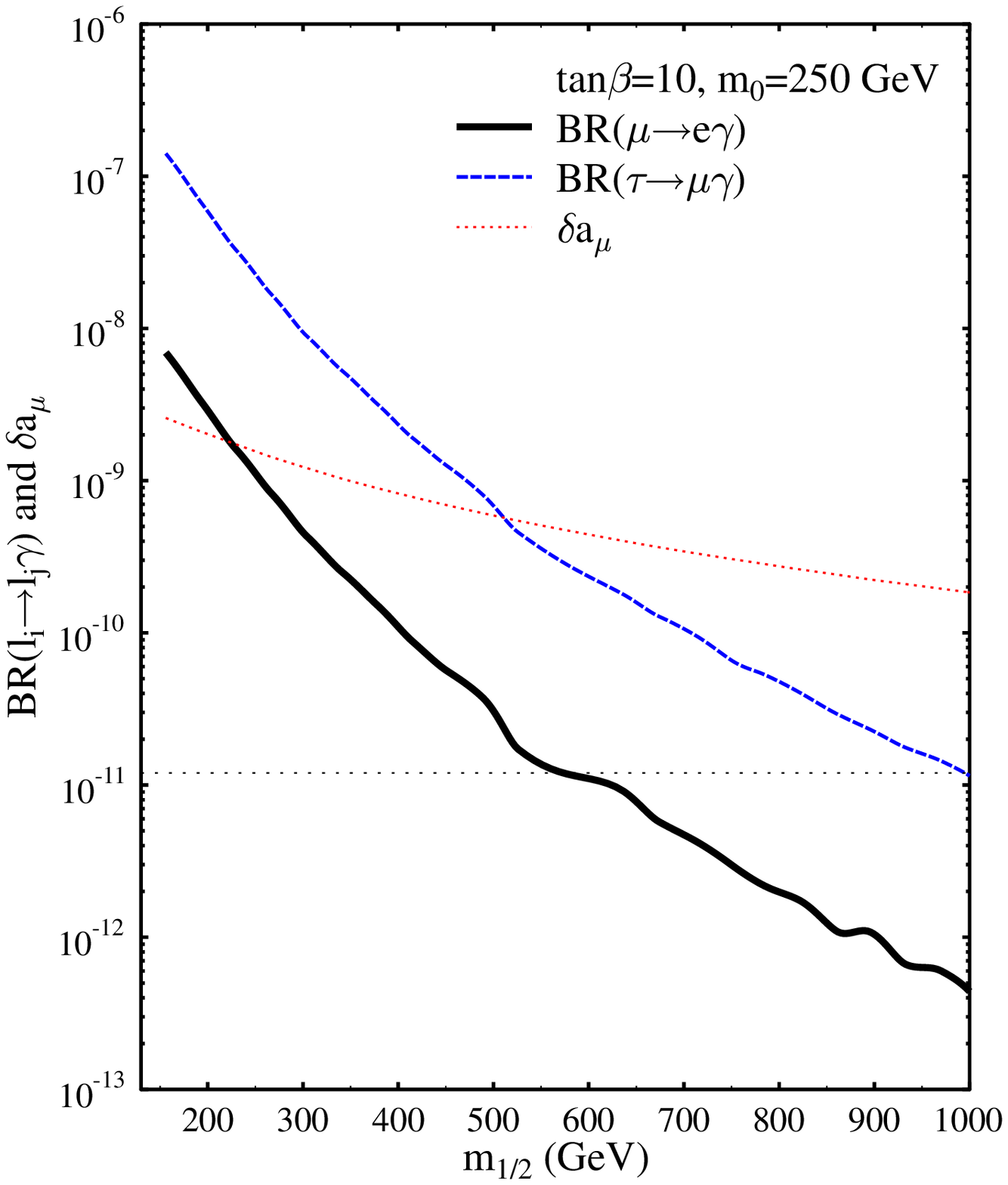}
\includegraphics[scale=0.5]{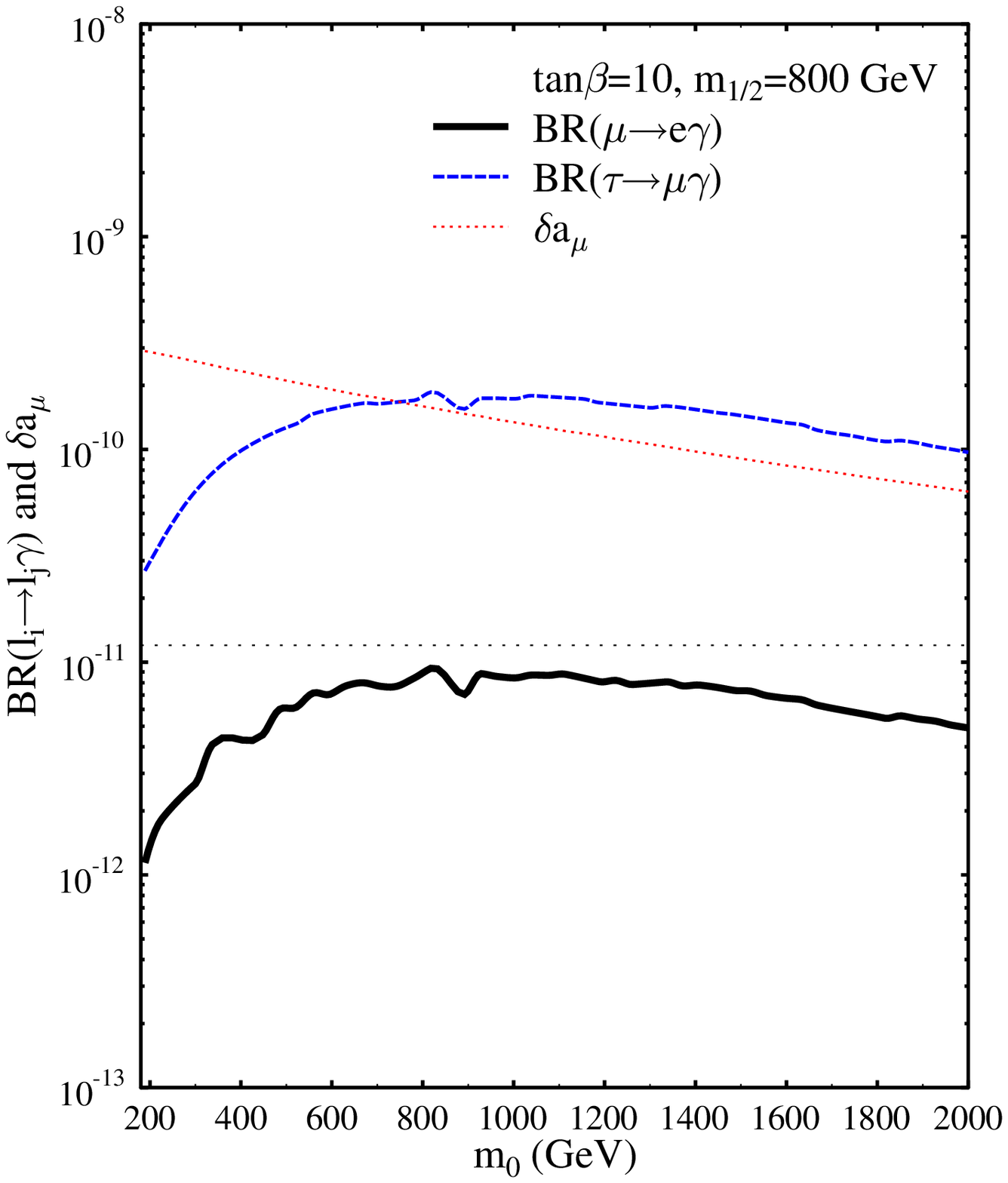}
\caption{\label{fig1}
BR($l_i\to l_j\gamma$) and
$\delta a_\mu$ as a function of $m_{1/2}$ and $m_0$ in
the left and right panels respectively.
$\tan\beta=10$, $A_0=0$ and $\mu>0$ are fixed.
We take $m_0=250 GeV$ for the left panel and $m_{1/2}=800 GeV$
for the right panel. The seesaw parameters are taken as
$\theta_T=\pi/4$, $\phi_i=0$, $\theta_{13}=0.05$
and $M_3=1\times 10^{15} GeV$. }
\end{figure}

In Fig. \ref{fig1}, we plot
BR($l_i\to l_j\gamma$) and $\delta a_\mu$
as functions of $m_{1/2}$ and $m_0$ for $\theta_T=\pi/4$
and $\phi_i=0$.  From this figure we can see that the process
$\mu\to e\gamma$ gives very strong constraint on the SUSY parameter
space: only with large $m_{1/2}$ and relatively small $m_0$
can its branching ratio be below the present experimental limit,
$1.2\times 10^{-11}$.
%It is very interesting to notice that similar conclusion
%is drawn when consider the constraint coming from the
%relic density of cold dark matter\cite{dm}.
For the following discussions,
we will fix $(m_{1/2},m_0)=(800, 250) GeV$. Since the muon anomalous
magnetic moment, $\delta a_\mu$, is nearly independent of the seesaw
parameters\cite{bi}, it is also fixed at about $2.7\times 10^{-10}$,
which will be omitted in the other figures.

\begin{figure}
\includegraphics[scale=0.8]{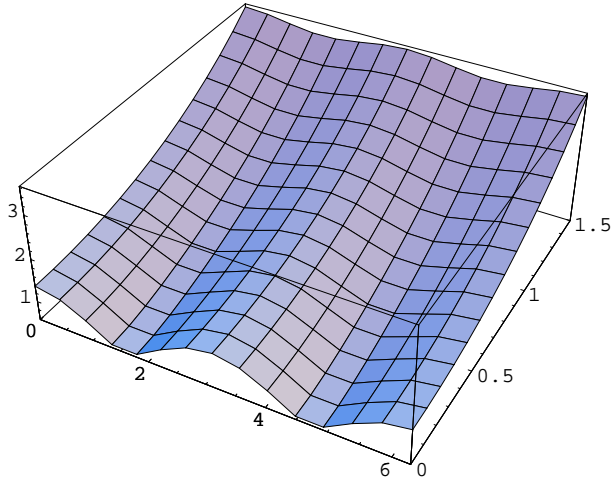}
\caption{\label{fig2}
$Y_3$ as function of $\theta_T$ for Re$\theta_T = 0 - 2\pi$
and Im$\theta_T = 0 - 1.5$. }
\end{figure}

Taking determinant on Eq. (\ref{2ss}) we know that the product of
$Y_{2,3}$ is fixed by the left- and right-handed Majorana neutrino masses.
The ratio of the two Yukawa couplings is determined by $\theta_T$.
In Fig. \ref{fig2} we show $Y_3$ as function of Re$\theta_T$
and Im$\theta_T$. Both the real and imaginary part influence
the ratio between $Y_2$ and $Y_3$. Since $Y_3$ increases almost
linearly with Im$\theta_T$, we expect BR($l_i\to l_j\gamma$)
also increase with Im$\theta_T$.

\begin{figure}
\includegraphics[scale=0.5]{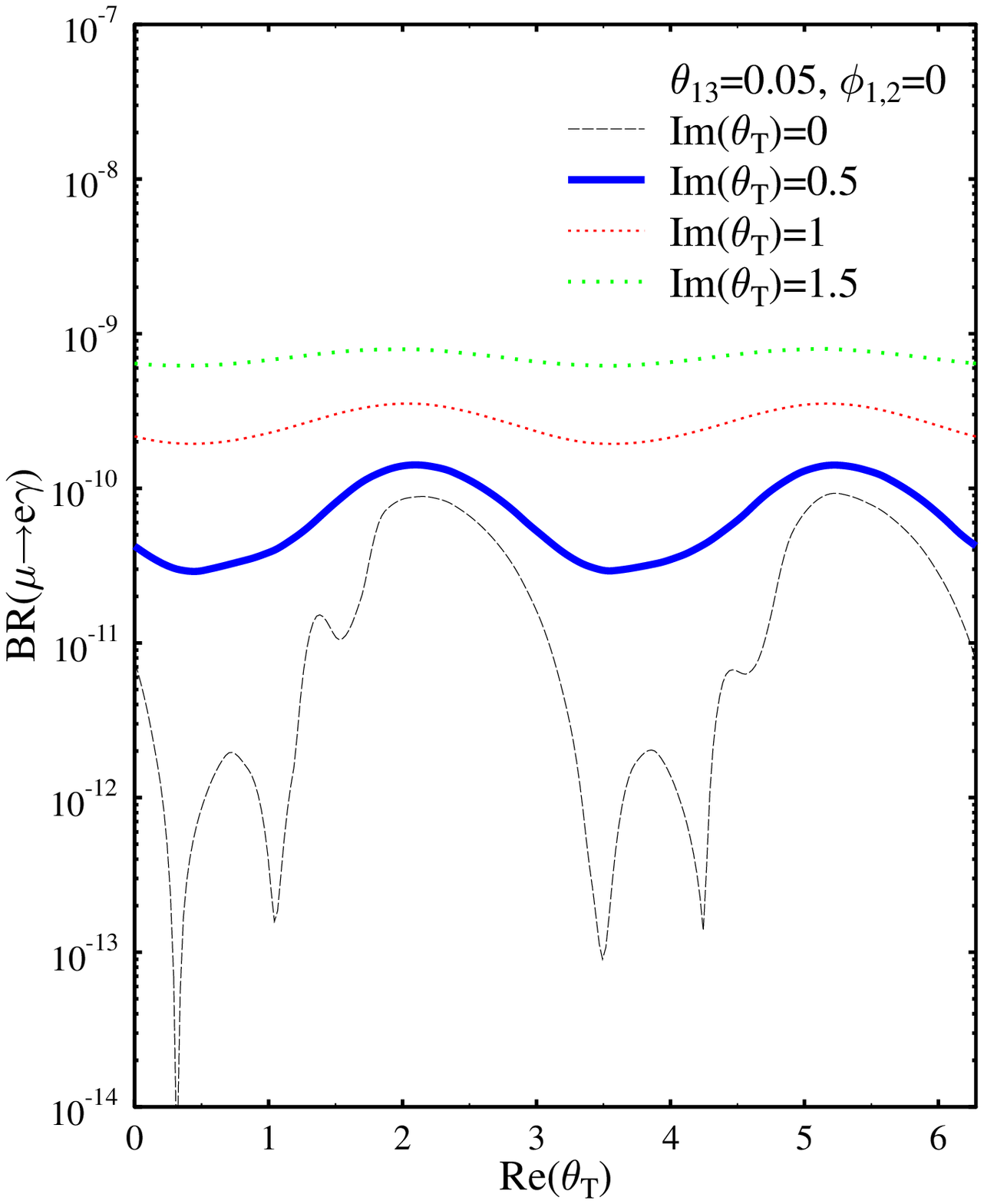}
\includegraphics[scale=0.5]{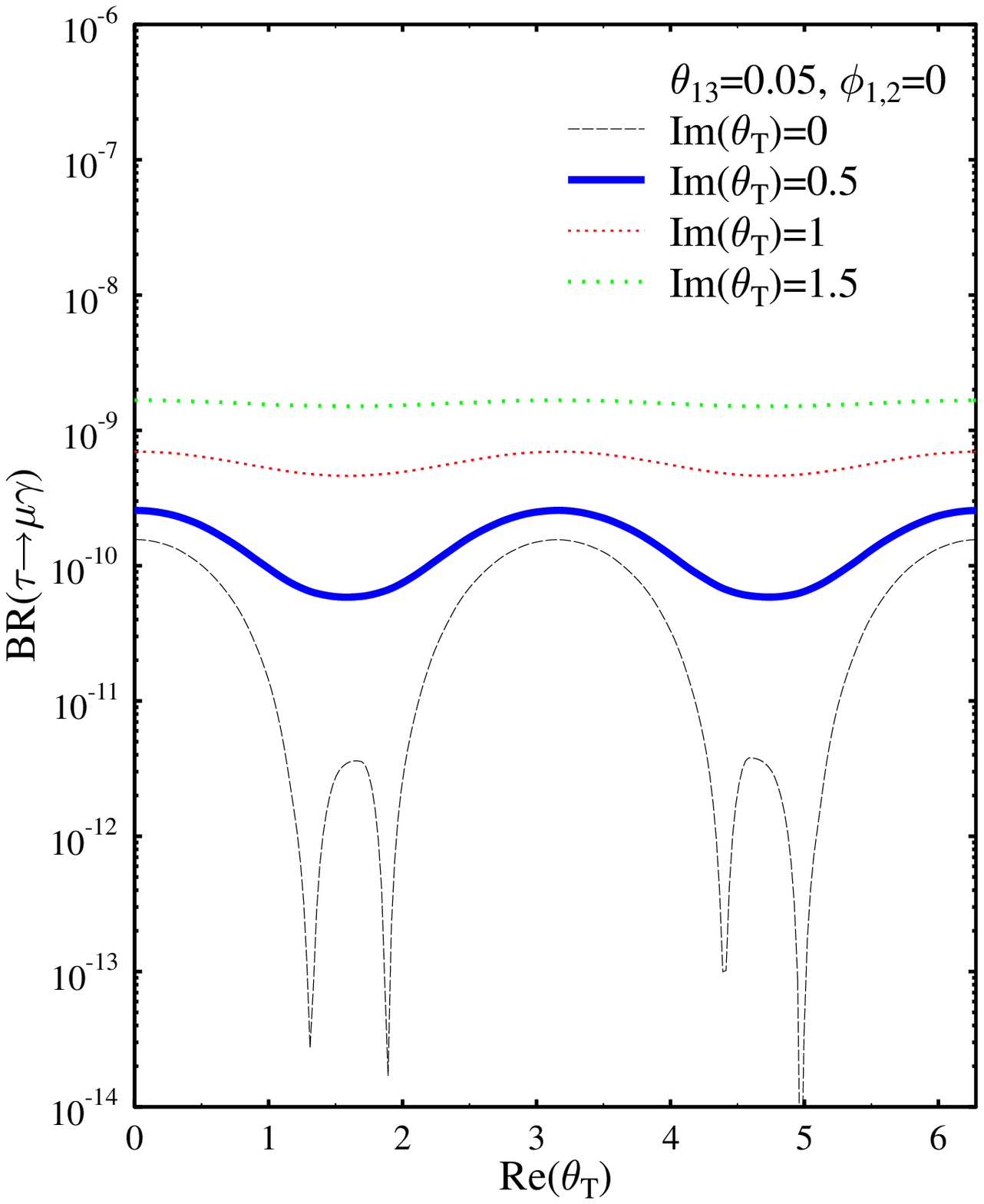}
\caption{\label{fig3}
BR($\mu\to e\gamma$) (Left) and BR($\tau\to \mu\gamma$) (Right)
as function of Re$\theta_T$ for Im$\theta_T = 0, 0.5, 1, 1.5$.
We fix $\theta_{13}=0.05$, $\phi_i=0$ and $M_3=1\times 10^{15} GeV$.
}
\end{figure}

In Fig. \ref{fig3}, we plot BR($\mu\to e\gamma$) and
BR($\tau\to \mu\gamma$) as function of
Re$\theta_T$ on the left and right panels respectively.
For Im${\theta}_T = 0.5$, BR($\mu\to e\gamma$) has been
greater than the experimental limit.

\begin{figure}
\includegraphics[scale=0.6]{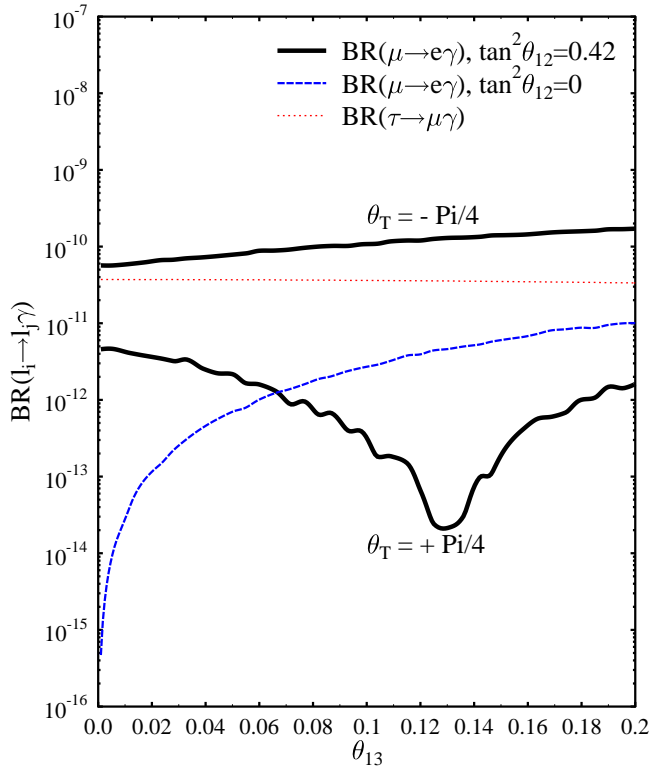}
\caption{\label{fig4}
BR($l_i\to l_j\gamma$) as function of $\theta_{13}$.
We fix $\phi_i=0$ and $M_3=1\times 10^{15} GeV$.
}
\end{figure}

In Fig. \ref{fig4}, BR($l_i\to l_j\gamma$) is drawn as function
of $\theta_{13}$. We can see BR($\mu\to e\gamma$) is very sensitive to
$\theta_{13}$, while BR($\tau\to \mu\gamma$) is insensitive
to $\theta_{13}$. The behavior in this figure is understood
if we notice that the flavor mixing between the first and
the second generations is nearly proportional to $V_{13}V_{23}^*Y_3^2$,
where $V_{13}=(U_\nu)_{12}V'_{23}+(U_\nu)_{13}V'_{33}$.
The two terms are added constructively or destructively, depending
on the sign of $\theta_T$. When we set $\theta_{12}=0$,
the branching ratio of $\mu\to e\gamma$ increases rapidly
with $\theta_{13}$, independent of the value of $\theta_T$.

\begin{figure}
\includegraphics[scale=0.6]{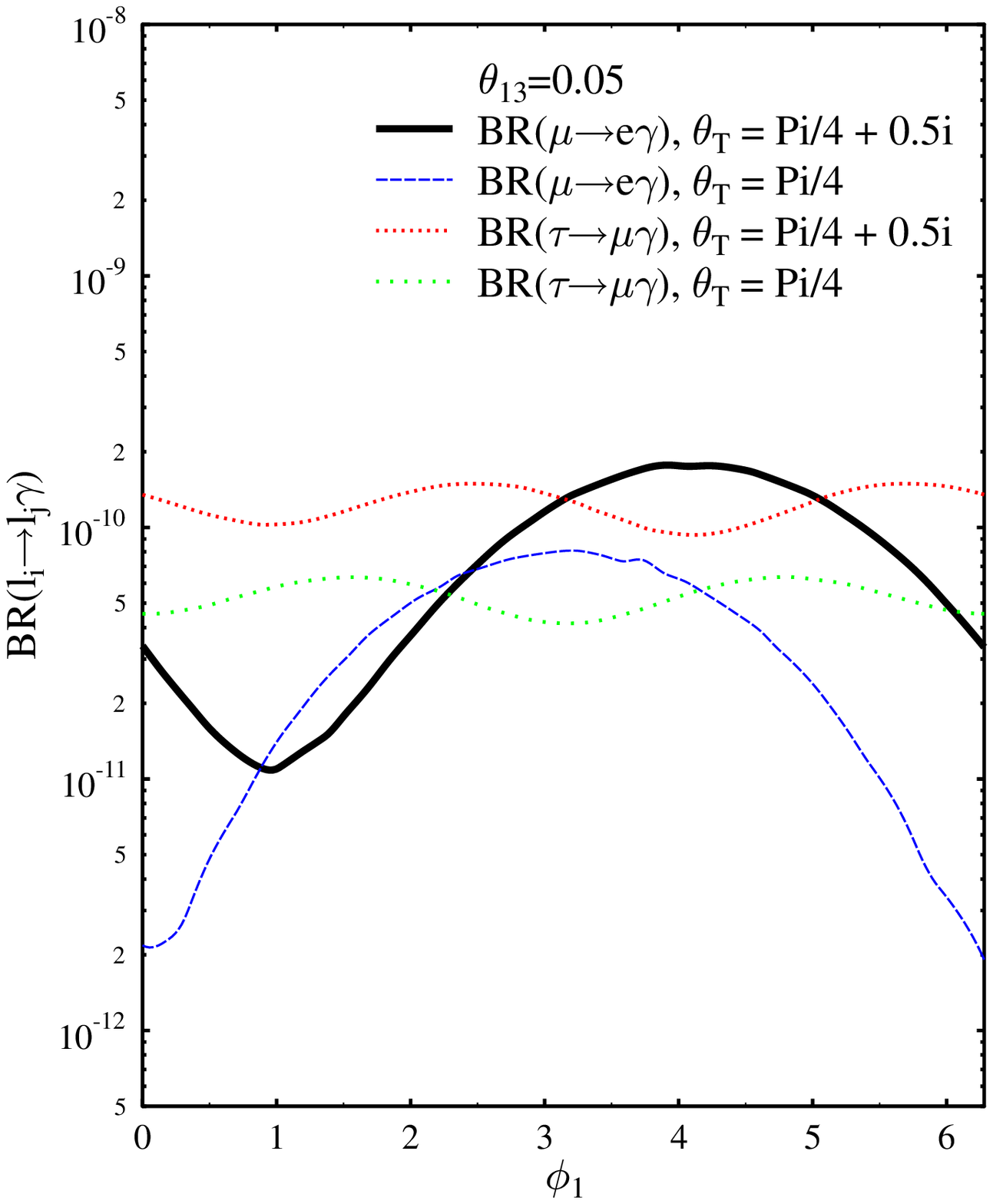}
\caption{\label{fig5}
BR($l_i\to l_j\gamma$) as function of $\phi_1$.
We fix $\theta_{13}=0.05$, $\phi_2=0$ and $M_3=1\times 10^{15} GeV$.
}
\end{figure}

In Fig. \ref{fig5}, we plot BR($l_i\to l_j\gamma$) as
function of $\phi_1$, which determines
the relative phase between $U_\nu$ and $V'$.
The behavior in the figure is easy to understand.
We also examined that BR($l_i\to l_j\gamma$)
is indeed independent of $\phi_2$, as we expected.

\begin{figure}
\includegraphics[scale=0.6]{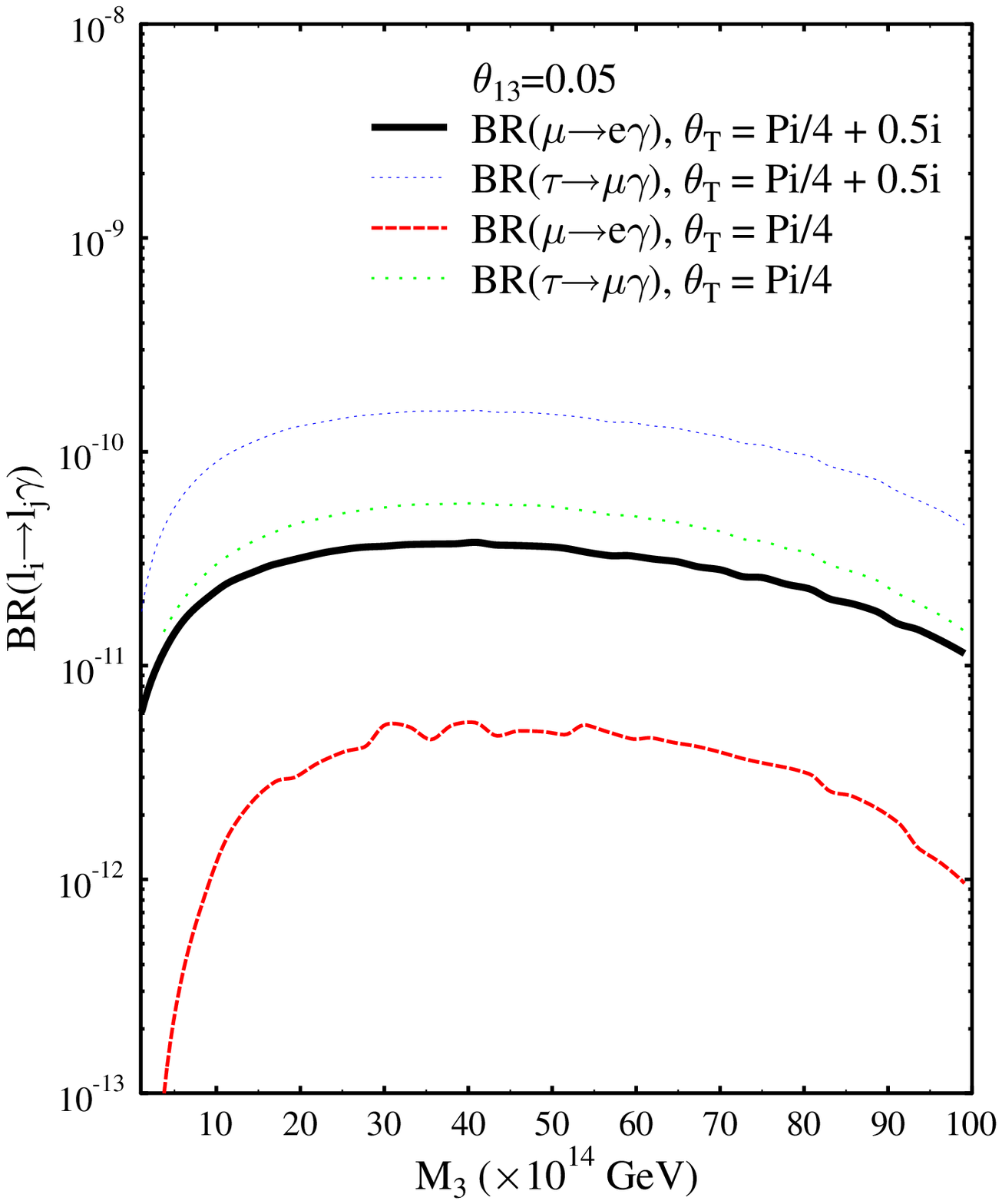}
\caption{\label{fig6}
BR($l_i\to l_j\gamma$) as function of $M_3$ for $\theta_T=\pi/4+0.5i$
and $\theta_T=\pi/4$.
We fix $\theta_{13}=0.05$ and $\phi_i=0$.
}
\end{figure}

Finally, we  plot BR($l_i\to l_j\gamma$) as function of
$M_3$. BR($l_i\to l_j\gamma$)
increases with $M_3$ at first, because it makes $Y_{2,3}$ larger.
However, when $M_3$ is as large as $10^{16} GeV$,
which is too close to $M_{GUT}$, the integration distance
$\log\frac{M_{GUT}}{M_3}$ becomes too small and the
branching ratio decreases. Although below $M_3$
LFV is still produced, see Eq. (\ref{ydy23}), the effects
are small, since the contribution from $Y_2^2$
is small, due to $Y_2^2 << Y_3^2$.
The $Y_3$ coupling contributes to the LFV below $M_3$ through
the mixing, $Y_3^2|\tilde{X}_{23}|^2$, which is also small
due to the small mixing element.

We have omitted BR($\tau\to e\gamma$) in all the figures
because the predicted branching ratio is much smaller than
the present experimental limit.

\subsection{Neutrinoless double beta decay}

From the above discussion we know that it is impossible to
produce degenerate solution for the left-handed neutrino
masses in this model. It is easy to estimate that
$<m>_{ee} = (2\sim 4)\times 10^{-3} eV$, depending on the value
of $U_{e3}$. So, in this sneutrino-inflaton model, it is
hard to account for the neutrinoless double beta
decay experimental signal\cite{0nubeta}.

\section{summary and discussions}

We have considered a double-sneutrino inflation model within the
minimal supersymmetric seesaw model.  With the mass ratio
$6\lesssim r \lesssim 10$ and the lighter sneutrino $M_1\sim 1.7
\times 10^{13}$ GeV, the model predicts a suppressed primordial
scalar spectrum around the largest scales which is favored at
$>1.5\sigma$.  The predicted CMB TT quadrupole is much better
suppressed than the single sneutrino model and the preference
level by the WMAP first year data is about $1.5\sigma$. Double quartic
inflation can also work very well in light of WMAP observations.

We then have studied the phenomenological implications of
this model. The seesaw parameters are constrained
by both particle physics and cosmological observations.
The strongest constraint comes from the required reheating
temperature by the gravitino problem. To some extend,
fine tunning is needed to
satisfy this constraint, which means that the right-handed
mixing angles $\theta_1$ and $\theta_2$ are much smaller than
the mass hierarchy of the right-handed neutrinos. Further,
the mass of the lightest left-handed neutrino should be at
the order of $10^{-10} eV$,
much smaller than the other two light neutrinos.

Leptogenesis arises from the decays of the cold inflaton--- the
lightest sneutrino. It is easy to account for the observed quantity of
the baryon number asymmetry in the Universe by
adjusting the seesaw parameters.
%However, since there is no direct relation between leptogenesis and
%other low energy phenomena, this adjustment for leptogenesis
%can not give definite predictions at low energies.

This model gives definite predictions on the lepton flavor
violating decay rates. In most parameter space, the branching ratio
of $\mu\to e\gamma$ is near or exceeds the present experimental
limit. However, the branching ratio of
$\tau\to \mu\gamma$ is at the order of about
$10^{-10} - 10^{-9}$, which is far below the current
experimental limit. Furthermore, in the appropriate range
of SUSY parameter space where LFV constraints are satisfied, 
the SUSY can only enhance
the muon anomalous magnetic moment at the amount of
$(2\sim 3)\times 10^{-10}$.

This model can not predict a degenerate light neutrino spectrum.
The observed signal of neutrinoless double beta decay,
if finally verified, can not be explained by the effective Majorana
neutrino mass in this model.

%Finally, we consider an alternative scenario
%with $M_3 \ll M_1$, say, $M_3 = 10^8 GeV$ (We still
%call this light sneutrino $\tilde{N}_3$). In this case the Eqs. (\ref{ydy11})
%and (\ref{mnd1}) are still correct. We again get the
%conclusion  that $m_{\nu_1} \approx 10^{-10} eV$ and
%$\hat{O}_{12}$, $\hat{O}_{13}$ are negligibly small.
%So, we can arrive at the formula Eq. (\ref{2ss}). With
%$Y_1\sim 10^{-5}$ and $Y_2\cdot Y_3\sim 10^{-4}$, it is
%hard to arrange the values of $Y_2$ and $Y_3$ in the same
%hierarchy as that of $M_i$.
%So, we can not avoid the `fine tunning' mentioned before by choosing
%$\tilde{N}_3$ much lighter.
%even if we choose $Y_3 \sim Y_1 \sim 10^{-5}$,
%$Y_2$ tends to unperturbatively large. If we choose both
%$Y_3$ and $Y_2$ at the order of $10^{-2}$, it becomes quite
%strange that no matter the heavier or lighter sneutrino
%has a larger Yukawa coupling than $\tilde{N}_1$. So, we
%conclude that this is not an attractive scenario.

\begin{acknowledgments}
We thank Z. Z. Xing for helpful discussions. We acknowledge the
using of CMBFAST program\cite{cmbfast,IEcmbfast}.  This work is
supported by the National Natural Science Foundation of China
under the grant No. 10105004, 19925523, 10047004
and also by the Ministry of Science and Technology of China under
grant No. NKBRSF G19990754.
\end{acknowledgments}

\newcommand\PRPT[3]{~Phys.Rept.{\bf ~#1}, #2~(#3)}

\end{document}